\pdfoutput=1

\documentclass[10pt,floatfix,hidelinks,twocolumn,aps,pre,groupedaddress]{revtex4-1}

\usepackage[caption=false]{subfig}
\usepackage{amsmath}
\usepackage{amssymb}
\usepackage{breqn}
\usepackage{dcolumn}
\usepackage{graphicx}
\usepackage{hyperref} 
\usepackage{standalone}
\usepackage{tikz} 
\definecolor{rubblue}{HTML}{003E64}

\newcolumntype{.}{D{.}{.}{-1}}
\newcommand{\mat}[1]{\boldsymbol{#1}}
\renewcommand{\vec}[1]{\boldsymbol{\mathbf{#1}}}
\definecolor{darkgreen}{rgb}{0,0.585,0}

\begin{document}


\title{Multi-Phase-Field Model for Surface/Phase-Boundary Diffusion}

\author{Raphael Schiedung}

\email{raphael.schiedung@rub.de}

\author{Reza Darvishi Kamachali}

\author{Ingo Steinbach}

\author{Fathollah Varnik}
\email{fathollah.varnik@rub.de}

\affiliation{Ruhr-Universit\"at Bochum \\
Interdisciplinary Centre for Advanced Materials Simulation (ICAMS)\\
Universit\"atsstr.~150$,$ 44801 Bochum$,$ Germany}

\date{\today}

\begin{abstract}
    The multi-phase-field approach is generalized to treat capillarity-driven diffusion parallel to the surfaces and phase-boundaries, i.e.\ the boundaries between a condensed phase and its vapor and the boundaries between two or multiple condensed phases.
    The effect of capillarity is modeled via curvature-dependence of the chemical potential whose gradient gives rise to diffusion.
    The model is used to study thermal grooving on the surface of a polycrystalline body.
    Decaying oscillations of the surface profile during thermal grooving, postulated by Hillert long ago but reported only in few studies so far, are observed and discussed.
    Furthermore, annealing of multi-nano-clusters on a deformable free surface is investigated using the proposed model.
    Results of these simulations suggest that the characteristic crater-like structure with an elevated perimeter, observed in recent experiments, is a transient non-equilibrium state during the annealing process.
\end{abstract}



\maketitle

\section{Introduction}
\label{Sec_Intro}
Diffusion in materials results from random motion of their constituents (atoms/molecules/voids).
Aside from self-diffusion of atoms or molecules within a homogeneous body in thermal equilibrium, diffusion is one of the most frequent transport mechanisms driven by weak deviations from the equilibrium state, such as a gradient in concentration or, more generally, chemical potential.
This latter aspect becomes particularly relevant at surfaces/interfaces with spatially varying curvature, which, due to Gibbs-Thompson relation, exhibit a gradient in chemical potential along the surface/phase boundary.
It is noteworthy that, while the capillarity driven diffusion at free surfaces is rather well investigated
(see e.g.~\cite{herring_surface_1999,mullins_theory_1957,gugenberger_comparison_2008}), much less attention has been payed to diffusion at the boundaries between two condensed phases.
This work focuses on this issue using the phase-field method.
In order to distinguish this type of diffusive transport from grain-boundary diffusion, in which boundaries are faster tracks for diffusion of solute atoms, the term `phase-boundary diffusion' is introduced here to express the capillarity-driven diffusion in boundaries.
A first investigation of a conserved phase-field model has been carried out by Caginalp in~\cite{caginalp_conserved-phase_1988,caginalp_dynamics_1990}.
Later, a conserved phase-field method has been used to simulate the evolution of elastically stressed films~\cite{ratz_surface_2006, yeon_phase_2006}.
These models, however, are limited to surface diffusion between two phases.
A comparison of the available phase-field models for surface diffusion is presented in~\cite{gugenberger_comparison_2008}.
In multiphase multi-grain systems, diffusion in phase-boundaries requires a treatment of multiple conserved phases at triple and higher-order junctions.

In this work, we present a conserved multi-phase-field method which is able to describe the capillarity-driven surface/phase-boundary diffusion in such complex topologies, where multiple junctions may be present.
The current method is based on the works reviewed by Gugenberger et al.~\cite{gugenberger_comparison_2008} on the two-phase surface diffusion models and the non-conserved multi-phase-field model \cite{steinbach_phase-field_2009,steinbach_phase-field_2013}.
For the sake of simplicity, here we consider immiscible phases without bulk diffusion and phase transformation, and solely focus on the effects of surface and phase-boundary diffusion.
The present approach can be used to investigate complex systems for which it is not possible to obtain an analytic solution or which are simply too large to be investigated by atomistic methods.
One of these possible scenarios is the solid-phase sintering of many particles where we can reach beyond the time and length scales of atomistic simulations.
As an example, we apply our model to study surface and phase-boundary diffusion of nano-clusters during annealing on a free surface.
This is similar to the experimental work presented in~\cite{kohler_thermally_2013}.
Furthermore, the construction of the current model allows the combination of surface/boundary diffusion with dissipative interface diffusion~\cite{steinbach_phase-field_2012,zhang_phase-field_2012}, elasticity~\cite{steinbach_multi_2006,mosler_novel_2014} and chemo-mechanical coupling~\cite{kamachali_dft-supported_2014,darvishi_kamachali_inverse_2017} effect in the same way that they are applied in non-conserved multi-phase-field methods.

The paper is organized as follows.
In Sec.~\ref{sec:thermo}, a short introduction into the thermodynamics of phase-boundary diffusion is given.
The equilibrium solution at triple junctions is discussed in Secs.~\ref{sec:not_deformable} and~\ref{sec:intro_neumann}.
The phase-boundary diffusion model is presented in Sec.~\ref{sec:model}.
Section~\ref{sec:results} compiles the obtained simulation results.
After a brief discussion of the equilibrium interface profile in Sec.~\ref{sec:results_profile}, thermal grooving is addressed in Sec.~\ref{sec:results_grooving}.
The late time profiles obtained from these simulations are then used in Sec.~\ref{sec:neumann} to check whether the method correctly recovers the von Neumann's triangle relation.
Section~\ref{sec:results_quadruple} highlights that the equilibrium behavior at the junctions is strongly dependent on the conservation constraints.
The evolution of multiple nano-clusters on a deformable surface is discussed in Sec.~\ref{sec:results_droplets}.
A summary compiles the main findings in Sec.~\ref{sec:summary}.
\subsection[Thermodynamics]{Thermodynamics of Surface Diffusion}
\label{sec:thermo}

We follow an approach similar to Mullins~\cite{mullins_theory_1957} and consider a body of single species surrounded by its vapor phase at constant temperature.
The surface is assumed to be in equilibrium with a finite curvature, $\kappa$, and the interface energy, $\sigma$.
The curvature is here considered as the sum of the principal curvatures, $\kappa_1$ and $\kappa_2$, with $\kappa = \kappa_1 + \kappa_2$.
The vapor pressure, $p_V$, is given by (assuming ideal gas and incompressible solid),
\begin{dmath} \label{eq:vapor_pressure}
    \ln \left( \frac{p_V}{p_0} \right)
    = \frac{\sigma V_m}{k_B T} \kappa
\end{dmath},
where $p_0$ is the vapor pressure for the planar surface, $k_B$ is the Boltzmann constant, $V_m$ is the molecular volume of the solid phase and $T$ is the temperature.
The chemical potential, $\mu$, of an ideal gas can be approximated as
\begin{dmath} \label{eq:mu_ideal_gas}
    \mu \approx - k_B T \ln \left( \frac{k_B T}{p_V \lambda^3} \right)
\end{dmath}.
In this relation, $\lambda$ is the thermal de Broglie wave length.
Inserting the vapor pressure, $p_V$, from Eq.~(\ref{eq:vapor_pressure}) into Eq.~(\ref{eq:mu_ideal_gas}) results in an expression for the dependency of the chemical potential on curvature,
\begin{dmath}
    \mu = - k_B T \ln \left( \frac{k_B T}{p_0 \lambda^3} \right)
            + \sigma V_m \kappa
\end{dmath}.
Considering a body of arbitrary curvature, this leads to a flux of particles, $\vec j$, from regions of high curvature to regions of low curvature,
\begin{dmath} \label{eq:surf_flux}
    \vec j
    = - \tilde{\mu} c \nabla_s \mu
    = - \tilde{\mu} \sigma c V_m \nabla_s \kappa
\end{dmath},
where $\tilde \mu$ is the mobility of the particles and $c$ the concentration of particles per unit area.
Here we treat the surface, $s$, as an isosurface of the concentration field with a constant value of $c$.
In a more general concept of multiple species, the interface energy may depend on the composition and thus the concentrations may vary alongside the surface as well.
Equation~(\ref{eq:surf_flux}) describes the flux of particles parallel to the surface.
The local increase of concentration of particles per unit area, $\dot c$, is obtained by taking the surface divergence of $\vec j$ so that
\begin{dmath} \label{eq:surface_diffusion}
    \dot c = \frac{D_s \sigma c V_m}{k_B T}\nabla_s^2 \kappa
\end{dmath},
where $D_s = \tilde \mu k_B T$ is the surface diffusion coefficient.
By multiplying $\dot c$ with the molecular volume, the surface normal velocity can be obtained,
\begin{dmath} \label{eq:vn_diffusion}
    v_n = \frac{D_s \sigma c V_m^2}{k_B T}\nabla_s^2 \kappa
\end{dmath}.
Equation~(\ref{eq:vn_diffusion}) is the basic equation which characterizes the dynamics of surface diffusion.
In non-conserved processes, e.g.\ phase transition, evaporation and condensation, the surface velocity is often described by a linear frictional ansatz
\begin{dmath} \label{eq:vn_eqvaporation}
    v_n = - \sigma M \kappa
\end{dmath},
where $M$ is a mobility coefficient (see~\cite{mullins_theory_1957}).
Usually, conserved surface diffusion and non-conserved phase transition processes are studied separately.
In principle, however, one can think of situations in which a combination of both effects determines the surface normal velocity,
\begin{dmath} \label{eq:vn}
    v_n
    = \sigma \left( \tilde M \nabla_s^2 \kappa - M \kappa \right)
\end{dmath},
in which a second mobility coefficient has been introduced, $ \tilde M$.
Equation (\ref{eq:vn}) states that the significance of the two contributions is determined by the coefficients $\tilde M$ and $M$, as well as the length scale of the process.
For smaller scales, i.e.\ larger curvatures, the effect of surface diffusion becomes increasingly more important.
It is to be noted that the dynamics of surface diffusion, Eq.~(\ref{eq:vn_diffusion}), for the solid/vapor interface is derived under the assumption of an ideal gas.
This is a good approximation for a capillary driven diffusion at a solid/vapor interface at elevated temperatures.

\subsection{Not Deformable Surface}
\label{sec:not_deformable}

A common example of three phases in contact is a liquid droplet, $\alpha$, on a flat non-deformable solid surface, $\beta$, surrounded by a gas phase, $\gamma$.
By minimizing the surface energy, one can obtain Young's law for the contact angle,
\begin{dmath} \label{eq:young}
    \cos{\theta} = \frac{\sigma_{\beta\gamma} - \sigma_{\beta\alpha}}{\sigma_{\gamma\alpha}}
\end{dmath}.
One can see that Eq.~(\ref{eq:young}) does not have a solution for all possible
values of the interface energies.
The droplet can either detach form the surface,
$ \frac{\sigma_{\beta\gamma} - \sigma_{\beta\alpha}}{\sigma_{\gamma\alpha}} \leq -1 $,
or can completely wet the surface,
$ \frac{\sigma_{\beta\gamma} - \sigma_{\beta\alpha}}{\sigma_{\gamma\alpha}} \geq 1 $.
More frequently, however, one has to deal with the case of partial wetting, described by the intermediate values, $ -1 < \frac{\sigma_{\beta\gamma} - \sigma_{\beta\alpha}}{\sigma_{\gamma\alpha}} < 1 $.

\begin{figure}
    \subfloat[]{\includestandalone[width=0.5\linewidth,mode=buildnew]{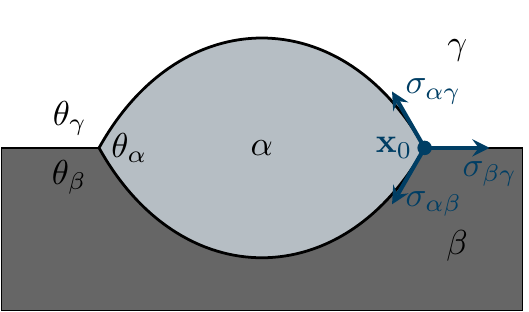}}
    \subfloat[\label{fig:neumann}]{\includestandalone[width=0.5\linewidth,mode=buildnew]{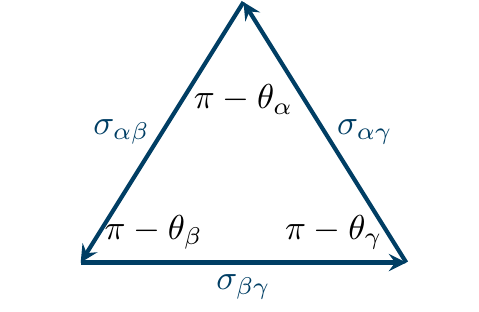}}
    \caption[Three coexisting mutual phases]{The figure (a) shows the three coexisting phases, $\alpha$, $\beta$ and $\gamma$ that are in contact to each other.
    $\theta_{\alpha}$, $\theta_{\beta}$ and $\theta_{\gamma}$ are the so called contact angles.
        $\sigma_{\alpha\beta}$, $\sigma_{\beta\gamma}$ and $ \sigma_{\alpha\gamma}$ are the interface energies acting on the triple point $\vec{x}_0$ alongside their interfaces.
        The von Neumann's triangle (b) visualizes the equilibrium condition, where the length of the triangle sides is given by the value of the interface energies.}
\end{figure}

\subsection{Deformable Surface}
\label{sec:intro_neumann}
The above consideration becomes more complex for three deformable phases in contact.
At a triple point in equilibrium, the sum of all forces acting on the point is zero and the force vectors form a triangle (see Fig.~\ref{fig:neumann}).
Then the law of sines can be applied to express the relation between the magnitude of the forces and their angles, $\theta_\alpha$, $\theta_\beta$ and $\theta_\gamma$, to each other,
\begin{equation} \label{eq:neumann_triangle}
    \frac{\sigma_{\beta \gamma}}{\sin{\theta_{\alpha}}}
    =
    \frac{\sigma_{\alpha \gamma}}{\sin{\theta_{\beta}}}
    =
    \frac{\sigma_{\alpha \beta}}{\sin{\theta_{\gamma}}}.
\end{equation}
This construction for the interface energies is referred to as the von Neumann's triangle relation (see e.g.~\cite{neumann_vorlesungen_1894,lester_contact_1961,shanahan_influence_1987}).
A comprehensive derivation of Eq.~(\ref{eq:neumann_triangle}) can be found in section 8.2 of the seminal textbook on molecular theory of capillarity by J.S.~Rowlinson and B.~Widom~\cite{rowlinson_molecular_1982}.
In the case of a junction between more than three phases, the situation becomes more complex.
The general approach, however, is the same and is based on the idea of interface energy minimization.

\section[The Model]{The Phase-Boundary Diffusion Model}
\label{sec:model}
In the following, we present a model for phase-boundary diffusion which is driven by the minimization of the local interface energy.
The present approach combines an existing non-conserved multi-phase-field model with the conserved phase-field models for surface diffusion at two-phase-boundaries.
The description of the referred multi-phase-field model is detailed in two articles~\cite{steinbach_phase-field_2009} and~\cite{steinbach_phase-field_2013}.
Furthermore, a review of surface diffusion models is given in~\cite{gugenberger_comparison_2008}.
In comparison to existing phase-field models for surface diffusion, the present model can describe the capillary driven diffusion of multiple phases.
Since the phase-boundary diffusion is the only process considered here, the volume of each individual phase is conserved.

\subsection[Interface Energy Functional]
{The Interface Free Energy Density Functional}
\label{sec:model_functional}

It is convenient to start with a suitable description of the free energy, $F$.
The free energy depends on the spacial configuration of $N$ phase-fields $\phi_\alpha \left( \vec{x} , t\right) $ with $\alpha \in \left[1,N\right]$, the spatial position $\vec{x}$ and the time $t$.
Each phase-field $\phi_\alpha$ indicates the location of a thermodynamic phase $\alpha$, in a way that $\phi_\alpha = 1$ means phase $\alpha$ is present and $\phi_\alpha = 0$ that it is not.
Additionally, a sum constrained for the phase-fields is used,
    $\sum_{\alpha=1}^N \phi_\alpha = 1$.
        In the case of phase-boundary diffusion, the phase-field value may be interpreted as an indicator function of a thermodynamic phase with a fixed composition and density.

\begin{figure}
    \includestandalone[width=\columnwidth, mode=buildnew]{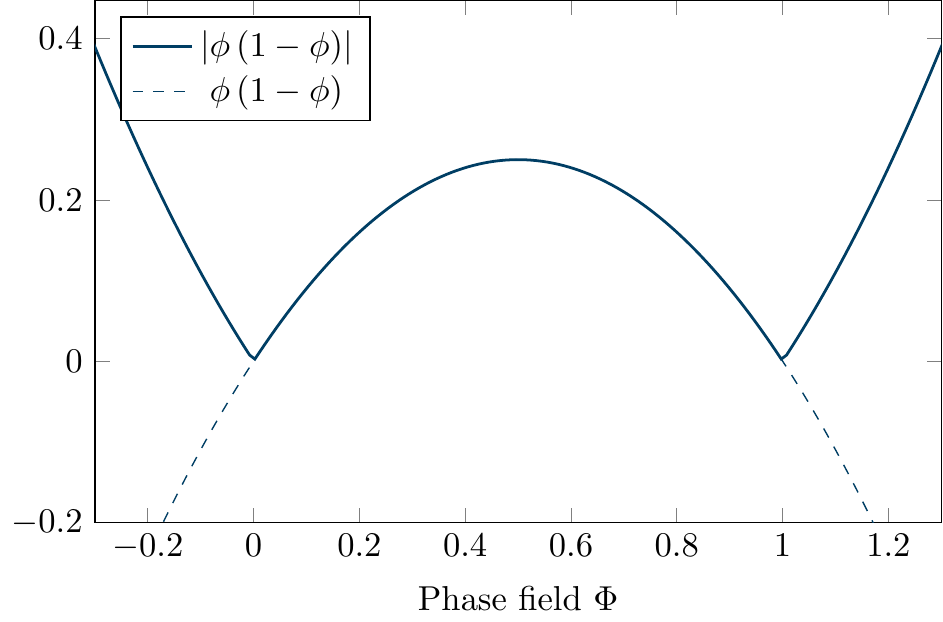}
    \caption{The potential function $\left\vert\phi_\alpha \phi_\beta\right\vert$ from Eq.~(\ref{eq:energy_density}) which reduces in the case of two phase-fields to $\left \vert \phi \left( 1 - \phi\right) \right \vert$ with $\phi = \phi_\alpha = 1 - \phi_\beta$.
    The dashed curve corresponds to $\phi \left( 1 -\phi \right)$.
    It has no minimum, but tends to $- \infty$ as $\phi \rightarrow \pm \infty$.}
    \label{fig:double_obstacle}
\end{figure}

Following the multi-phase-field approach in~\cite{steinbach_phase-field_2009, steinbach_phase-field_2013}, the total free energy can be written as a functional  $F\left(\left\{\phi_{\alpha} \right\}\right)$.
This free energy functional is defined through the free energy densities, $f_{\alpha \beta}$, of the interface between the phases $\alpha$ and $\beta$.
The sum of all interface energy densities is integrated over the volume of interest, $\Omega$,
\begin{dmath} \label{eq:energy_functional}
    F\left(\left\{\phi \right\}\right)
    = \sum^{N}_{\alpha=1} \sum^{N}_{\beta = \alpha +1}
    \int_\Omega f_{\alpha \beta}
    \left(\phi_{\alpha}, \phi_{\beta}\right)
    \mathrm \, dV
\end{dmath}.
One may add other energy densities into the concept, in order to include other physical effects.
The energy densities of the interfaces are proposed in a pair-wise manner between existing phase-fields,
\begin{dmath} \label{eq:energy_density}
    f_{\alpha \beta}
    =
    \frac{4\sigma_{\alpha\beta}}{\eta}
    \left(
        - \frac{\eta^2}{\pi^2}
        \nabla \phi_{\alpha} \cdot \nabla \phi_{\beta}
        +
        \left\vert \phi_{\alpha} \phi_{\beta} \right\vert
    \right)
\end{dmath}.
Here, $\eta$ defines the width of the transition region between two phase-fields, e.g., $\phi_\alpha$ and $\phi_\beta$.
$\left\vert \phi_{\alpha} \phi_{\beta} \right\vert$ is called double obstacle potential and restricts the value of the phase-field to the interval $\left[0,1\right]$.
Figure~\ref{fig:double_obstacle} shows a plot of the double obstacle potential which can be simplified to $\left\vert\phi \left(1- \phi\right)\right\vert$ with $\phi = \phi_\alpha = 1 - \phi_\beta$ in the case of only two phase-fields.
As it can be seen from Fig.~\ref{fig:double_obstacle}, the use of absolute value is necessary to ensure that $ \phi = 0$ and $ \phi = 1$ correspond to two equilibrium solutions.

\subsection[Phase-Boundary Dynamics]{The Dynamics of Phase-Boundary Diffusion}
\label{sec:model_dynamics}

The dynamic equations for phase-boundary diffusion can now be derived by using the variational derivative of the free energy functional with respect to the phase-field,
\begin{dmath} \label{eq:interface_diffusion}
    \dot{\phi}_{\alpha}
    = - \frac{1}{N} \sum_{\beta=1}^{N}
        \nabla \cdot \mat{\tilde M}_{\alpha \beta}
        \nabla \left( \frac{\delta}{\delta \phi_\alpha}
        - \frac{\delta}{\delta \phi_\beta} \right) F
    = - \frac{1}{N} \sum_{\beta=1}^{N}
        \nabla \cdot \mat{\tilde M}_{\alpha \beta} \nabla
        \psi_{\alpha \beta}
\end{dmath}.
Here, $\mat{\tilde M}_{\alpha \beta}=\mat{\tilde M}_{\beta\alpha} $ is a symmetric mobility tensor of the interface between the phases $\alpha$ and $\beta$.
    It is of fourth-rank with $\tilde M_{j i \alpha \beta} = \tilde M_{i j \alpha \beta}$, where $i$ and $j$ indicate Cartesian coordinates, $i,j \in \{x,y,z\}$.
Equation~(\ref{eq:interface_diffusion}) ensures that each phase-field, $\phi_\alpha$, is conserved.
Moreover, since $\sum^N_{\alpha=1} \dot \phi_\alpha = 0 $,  it follows that the sum rule, $\sum_{\alpha=1}^N \phi_\alpha(t) = 1$, is fulfilled at all times, $t$, provided that it is valid at $t=0$.
    In Eq.~(\ref{eq:interface_diffusion}), $\psi_{\alpha \beta}$ is a generalized diffusion potential, defined via
\begin{dmath} \label{eq:generalized_chemical_potential}
    \psi_{\alpha \beta}
    =\left( \frac{\delta}{\delta \phi_\alpha}
    - \frac{\delta}{\delta \phi_\beta} \right) F
    = I_{\alpha \beta} - I_{\beta \alpha}
    + \sum^N_{\gamma = 1, \gamma \neq \alpha, \gamma \neq \beta}
    \left(I_{\alpha \gamma} - I_{\beta \gamma}\right)
\end{dmath}.
The function $I_{\alpha \beta}$ is a shorthand notation for the functional derivative of the free interface density $f_{\alpha \beta}$,
\begin{dgroup}
    \begin{dmath}
        I_{\alpha \beta}
        =
        \frac{\partial f_{\alpha \beta}}{\partial \phi_\alpha}
        - \nabla \frac{\partial f_{\alpha \beta}}{\partial \nabla \phi_\alpha}
    \end{dmath}.
    \begin{dmath}
        =
        \frac{4 \sigma_{\alpha\beta} }{\eta}
        \left(
        \frac{\eta^2}{\pi^2}
        \nabla^2 \phi_\beta
        + \frac{\partial \left\vert \phi_\alpha \phi_\beta \right\vert}
        {\partial \phi_\alpha}
        \right)
        \label{eq:Iab_b}
    \end{dmath}.
    \label{eq:I_ab}
\end{dgroup}
It should be remarked here that $I_{\alpha \beta} \neq I_{\beta \alpha}$.
In general, the interface energy, $\sigma_{\alpha \beta} \left( \vec{n}_{\alpha \beta} \right)$, is a function of the interface orientation which can be determined by using the interface normal vector,
\begin{dmath} \label{eq:interface_normal}
    \vec{n}_{\alpha \beta}
    =
    \frac{\phi_\alpha \nabla \phi_\beta - \phi_\beta \nabla \phi_\alpha}
    {\left\Vert \phi_\alpha \nabla \phi_\beta
    - \phi_\beta \nabla \phi_\alpha \right\Vert}
\end{dmath}.
This leads to additional terms in Eq.~(\ref{eq:I_ab}) such as the Herring torque~\cite{herring_surface_1999}.
For the sake of simplicity, however, this study is restricted to the case of a constant interface energy.
A comment is necessary here regarding the use of a three dimensional $\nabla$ operator in Eq.~(\ref{eq:interface_diffusion}) instead of the surface gradient $\nabla_s$ (see Eq.~(\ref{eq:surface_diffusion})).
This is a consequence of the fact that, in diffuse interface approaches such as the present multi-phase-field model, a strictly two dimensional interface does not exist.
Rather, it is replaced by an interface layer with a finite thickness, $\eta$.
In the limit of a sharp interface $(\eta\rightarrow 0)$, however, Eq.~(\ref{eq:interface_diffusion}) approaches Eq.~(\ref{eq:surface_diffusion}).
This is ensured by an appropriate choice of the mobility tensor, $\mat{\tilde M}_{\alpha\beta}$, in such a way as to effectively restrict the diffusion flux to the directions tangential to the interface.
Diffusion along the normal direction is only allowed in order to keep the phase-field profile stable and vanishes in the limit of a sharp interface.
An asymptotic analysis of multiple surface diffusion models is performed in~\cite{gugenberger_comparison_2008}.
Here we set,
\begin{dmath} \label{eq:tensorial_mobility}
    \mat{\tilde M_{\alpha\beta}}
    =
    \tilde M_{\alpha\beta} \left( \mat{1} - a_{\alpha\beta} \vec{n}_{\alpha\beta}\vec{n}_{\alpha\beta} \right)
\end{dmath}.
In this way, the tensorial mobility, $\mat{\tilde M}_{\alpha\beta}$, is restricted to the tangential plane of the interface by using the interface normal $\vec{n}_{\alpha\beta}$.
In Eq.~(\ref{eq:tensorial_mobility}), $\tilde M_{\alpha\beta}$ is the scalar magnitude of $\mat{\tilde M}_{\alpha\beta}$ and $\mat{1}$ is the unit tensor.
Additionally, the function $a_{\alpha \beta}$ interpolates between a purely isotropic mobility ($a_{\alpha\beta} = 0$) and a pure tangential one ($a_{\alpha\beta} = 1$).
Thus, a flux across the interface is generated only if the interface is not in its equilibrium shape,
\begin{dmath} \label{eq:ratio}
    a_{\alpha \beta}
    =
    \operatorname{limit}_{1}
    \left(
        \frac{\eta^2}{\pi^2} \frac{\left\vert \nabla \phi_\alpha \cdot
         \nabla \phi_\beta \right\vert}{\left\vert
         \phi_\alpha \right\vert \left\vert\phi_\beta \right\vert}
    \right)
\end{dmath}.
The operator $\operatorname{limit}_{1}$ restricts the interpolation function $a_{\alpha\beta}$ to the interval $\left[0,1\right]$ with
\begin{equation}
    \operatorname{limit}_{1} \left( x \right)
      =
      \begin{cases}
          1 & x > 1 \\
            x & \text{else}.
      \end{cases}
\end{equation}
If the sharp interface limit is not of major interest, it is tempting to use the simpler isotropic mobility tensor,
\begin{dmath} \label{eq:scalar_mobility}
    \mat{\tilde{M}}_{\alpha \beta} = \tilde{M}_{\alpha \beta} \mat{1}
\end{dmath}.
However, it is shown in~\cite{gugenberger_comparison_2008} that an isotropic mobility tensor is less suitable to model the dynamics of surface diffusion.
Therefore, with the exception of a single test shown in Fig.~\ref{fig:profile_curved}, all the simulations reported here are performed using the non-isotropic mobility tensor, Eq.~(\ref{eq:tensorial_mobility}).

\subsection{Simulation Details}
\label{sec:model_procedure}
It is noteworthy that the derivatives of the obstacle potential, $\partial \left\vert \phi_\alpha \right\vert / \partial \phi_\alpha$ and $\partial \left\vert \phi_\beta \right\vert / \partial \phi_\beta$,
are not continuous at $ \phi = 0 $ and $ \phi = 1 $.
    Therefore, a so called bent-cable model~\cite{chiu_bent-cable_2002} is used in order to piecewise interpolate between the linear derivative of the double obstacle and the non-linear interpolation at the minima.

The model is implemented inside the open source software project OpenPhase (www.OpenPhase.de).
A finite difference scheme with a 27-point stencil~\cite{spotz_high-order_1995}
is used for the discretization of the Laplacian in Eq.~(\ref{eq:I_ab}).
This high number of stencil points is used in order to avoid numerical errors.
We also tested the method with a 7-point stencil and the results showed no significant difference here.
This observation is corroborated by similar studies with a focus on curvature evaluation~\cite{vakili_numerical_2017}.
The divergence and the gradient in Eq.~(\ref{eq:interface_diffusion}) are both discretized with a 3-point first order central finite difference scheme.
A first order explicit Euler method is used for the discretization of the time derivative.
Because phase-boundary diffusion becomes more significant on small length scales, a spacial discretization of $\Delta x = 10^{-9}$m and temporal discretization of $\Delta t = 10^{-13}$s is considered.
The surface/interface energies used in this study are of the order of $\sigma \sim 1$ J/m$^2$, which is typical for metallic systems.
The mobility coefficient of Eq.~(\ref{eq:interface_diffusion}) is set to $\tilde M_{\alpha \beta} = 10^{-25}\text{m}^2/\text{Js}$.
This value of the interface mobility does not correspond to a real physical system but is necessary to keep the algorithm stable.
Results obtained in this work are thus of generic rather than material specific nature.
To better reflect this feature, all lengths and times are reported below in units of $\Delta x$ and $\Delta t$.
If not otherwise stated, the interface width is set to $\eta = 10 \Delta x$.
In order to calculate the right hand side of Eq.~(\ref{eq:generalized_chemical_potential}), three additional boundary grid cells for $\phi$ are used around the computation domain.
One way to determine these boundary cells is to use periodic boundary conditions.
The volume of the phases is in this case conserved up to the machine's roundoff error.

\section{Results and Discussion}
\label{sec:results}
The phase-boundary diffusion model is first tested with regard to the equilibrium interface profile for a spherically symmetric phase-field in Sec.~\ref{sec:results_profile}.
A detailed study of thermal grooving is presented in Sec.~\ref{sec:results_grooving}.
Assuming that the late time profile obtained from these simulations is a good approximation for equilibrium shape of a three phase boundary, it is examined in Sec.~\ref{sec:neumann} whether these simulations are consistent with the von Neumann's triangle relation.
As an example for the influence of periodic boundaries, we simulate a 3D tessellation with octahedra in Sec.~\ref{sec:results_quadruple}.
In Sec.~\ref{sec:results_droplets}, the annealing of nano-clusters on a deformable, initially flat, surface is investigated.
The results obtained from these simulations are discussed in the context of the available literature.

\subsection{Interface Profile}
\label{sec:results_profile}

\begin{figure*}
    \subfloat[\label{fig:profile_curved_tensorial_mobility}]{\includestandalone[width=0.5\linewidth,mode=buildnew]{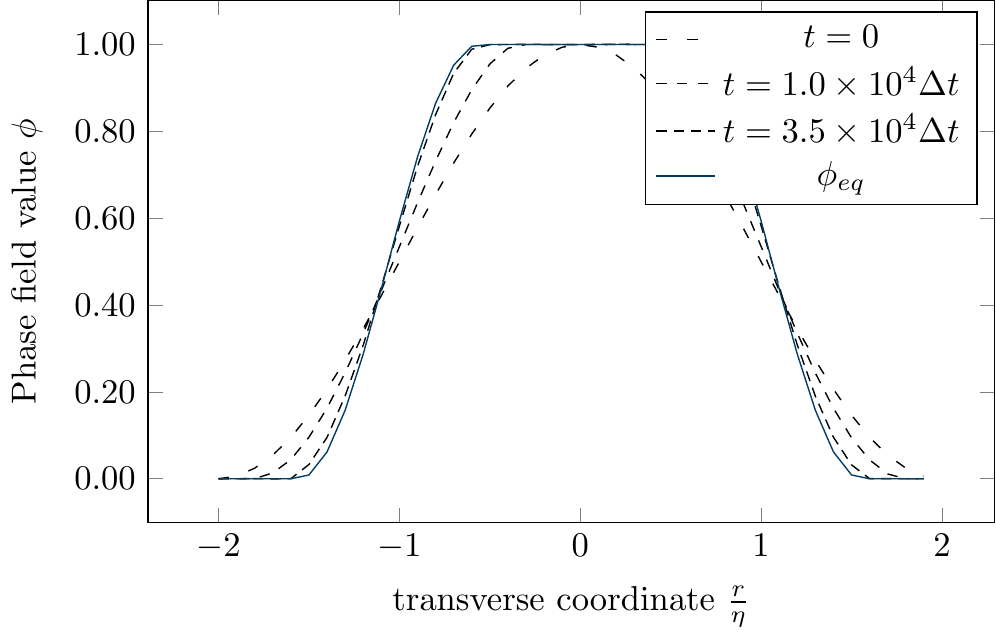}}
    \subfloat[\label{fig:profile_curved_scalar_mobility}]{\includestandalone[width=0.5\linewidth,mode=buildnew]{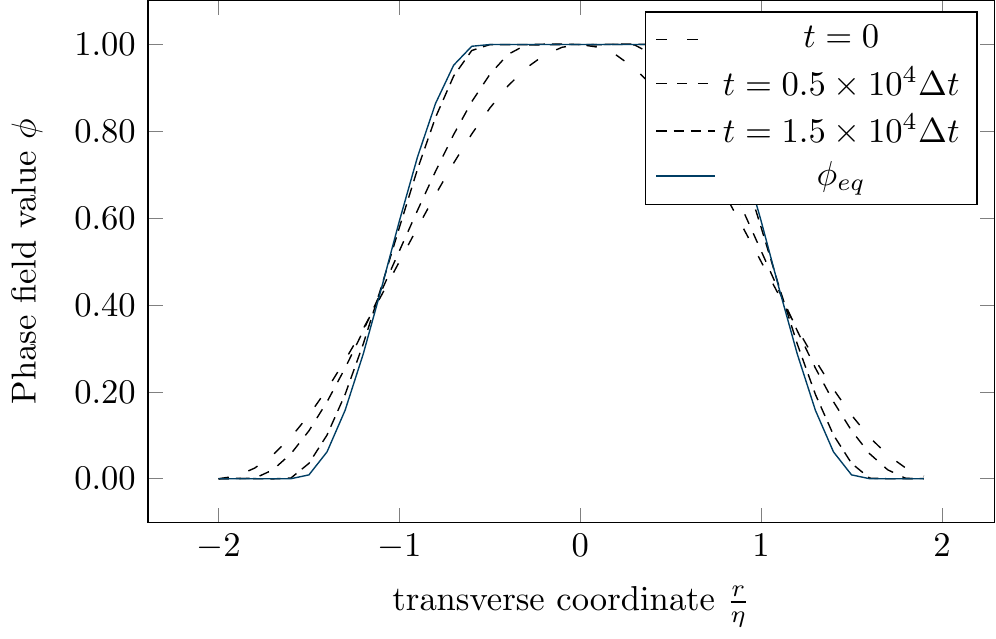}}
    \caption[Curved interface profile relaxation]{Relaxation of the interface profile, across the center of sphere, towards its equilibrium shape with (a) a non-isotropic and (b) an isotropic mobility tensor.
        For this study, a two dimensional disk with a radius of $R =  \eta$ and an initial interface width of $\eta_{\text{initial}} = 2 \eta$ has been initialized with $\eta = 10 \Delta x$ and $\Delta x = 10^{-9}$m (see Eq.~(\ref{eq:travel_sphere})).}
    \label{fig:profile_curved}
\end{figure*}

As a very first test, we consider relaxation of the interface towards its equilibrium profile, similar to the test performed in~\cite{gugenberger_comparison_2008}.
In the case of only two phases and a planar interface, the analytical solution of the interface profile is known  at equilibrium (see App.~\ref{app:travel}).
We show numerically that the phase-field profile obtained from the planar interface is also a good approximation for the profile of a spherically symmetric phase-field.
For two phase-fields with
$\phi_\alpha = 1 - \phi_\beta = \phi$,
    \begin{dmath} \label{eq:travel_sphere}
        \phi_{\text{eq}}
        =
        \begin{cases}
            1 & \text{if }  r < R - \frac{\eta}{2} \\
            \frac{1}{2} - \frac{1}{2} \sin\left(\frac{\pi \left(r-R\right)}{\eta}\right)
            &\text{if } R - \frac{\eta}{2} \leq r \leq R + \frac{\eta}{2}\\
            0 & \text{else},
        \end{cases}
    \end{dmath}
in which, $r$ is the radial coordinate and $R$ the radius of the sphere.

Equation~(\ref{eq:travel_sphere}) is used as initialization for the phase-field, but with a wider interface width ($\eta_{\text{initial}} = 2 \eta$).
Figure~\ref{fig:profile_curved} shows the value of $\phi$ along the center line of the sphere for different time steps.
In the current set-up, the phase-field at the initial time, $t=0$, is already in a spherical shape, but the interface is wider than the equilibrium solution.
Thus, diffusion is expected to occur only in the direction normal to the interface.
A survey of Fig.~\ref{fig:profile_curved} reveals that isotropic and non-isotropic tensorial mobilities result in different relaxation rates towards the equilibrium profile.
This is closely related to the fact that, in the case of a non-isotropic mobility tensor, the matrix-elements are chosen such that the transport is mainly allowed along the tangential direction, while it is rather restricted along the normal direction.
In the present set-up, where the driving force acts only along the normal direction, this leads to a slower diffusion as compared to an isotropic mobility tensor, where the diffusion process is not restricted along any direction.

\subsection[Thermal Grooving]{An Example of Thermal Grooving}
\label{sec:results_grooving}

Mullins has investigated the mechanisms of surface diffusion and evaporation-condensation which lead to the formation of surface grooves in a polycrystal that is heated up to elevated temperatures~\cite{mullins_theory_1957}.
An initially flat surface with a grain boundary perpendicular to the surface is considered.
For surface diffusion, a first order approximation of the surface profile evolution,
\begin{equation} \label{eq:mullins}
    y_{M} \left(x, t \right)
    = m (Bt)^{\frac{1}{4}}
    \sum_{n=0} a_n \left[\frac{x}{(Bt)^{\frac{1}{4}}}\right]^n
    + \mathcal{O} \left(m^2\right)\, ,
\end{equation}
has been obtained for small slopes $m$ and with the parameter $B = \tilde \mu \sigma c V_m^2$ (see~\cite{mullins_theory_1957}).
    The values for $a_n$ are given in Tab.~\ref{coefficients}.
The profile of Eq.~(\ref{eq:mullins}) is shape invariant and its amplitude, $Bt$, can not be changed without stretching it along the $x$-axis.
This solution suggests that the groove profile increases its amplitude over time without changing its shape.

\begin{ruledtabular}
    \begin{table}
        \begin{tabular}{c c l D{,}{.}{3} l c c}
            &&$a_0$    & -7,803 & $\times 10^{-1}$  &&\\
            &&$a_1$    & 10,000 & $\times 10^{-1}$  &&\\
            &&$a_2$    & -2,886 & $\times 10^{-1}$  &&\\
            &&$a_4$    &  8,130 & $\times 10^{-3}$  &&\\
            &&$a_6$    & -2,004 & $\times 10^{-4}$  &&\\
            &&$a_8$    &  3,625 & $\times 10^{-6}$  &&\\
            &&$a_{10}$ & -4,969 & $\times 10^{-8}$  &&\\
            &&$a_{12}$ &  5,339 & $\times 10^{-10}$ &&\\
            &&$a_{14}$ & -4,655 & $\times 10^{-12}$ &&\\
        \end{tabular}
        \caption{Polynomial coefficients of Eq.~(\ref{eq:mullins}).}
        \label{coefficients}
    \end{table}
\end{ruledtabular}

\begin{figure*}
    \subfloat[] {\includegraphics[width = 0.5\linewidth]{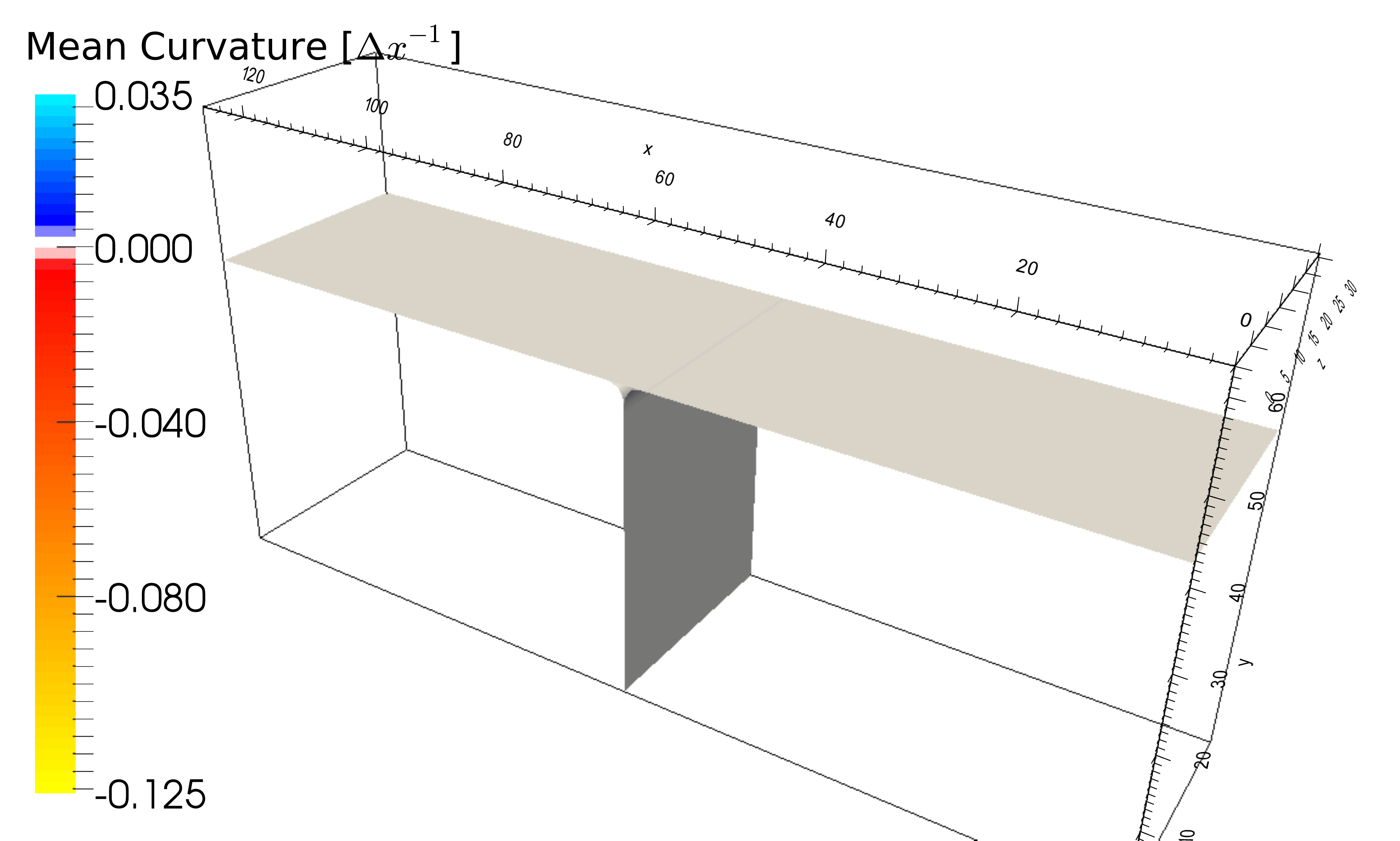}}
    \subfloat[] {\includegraphics[width = 0.5\linewidth]{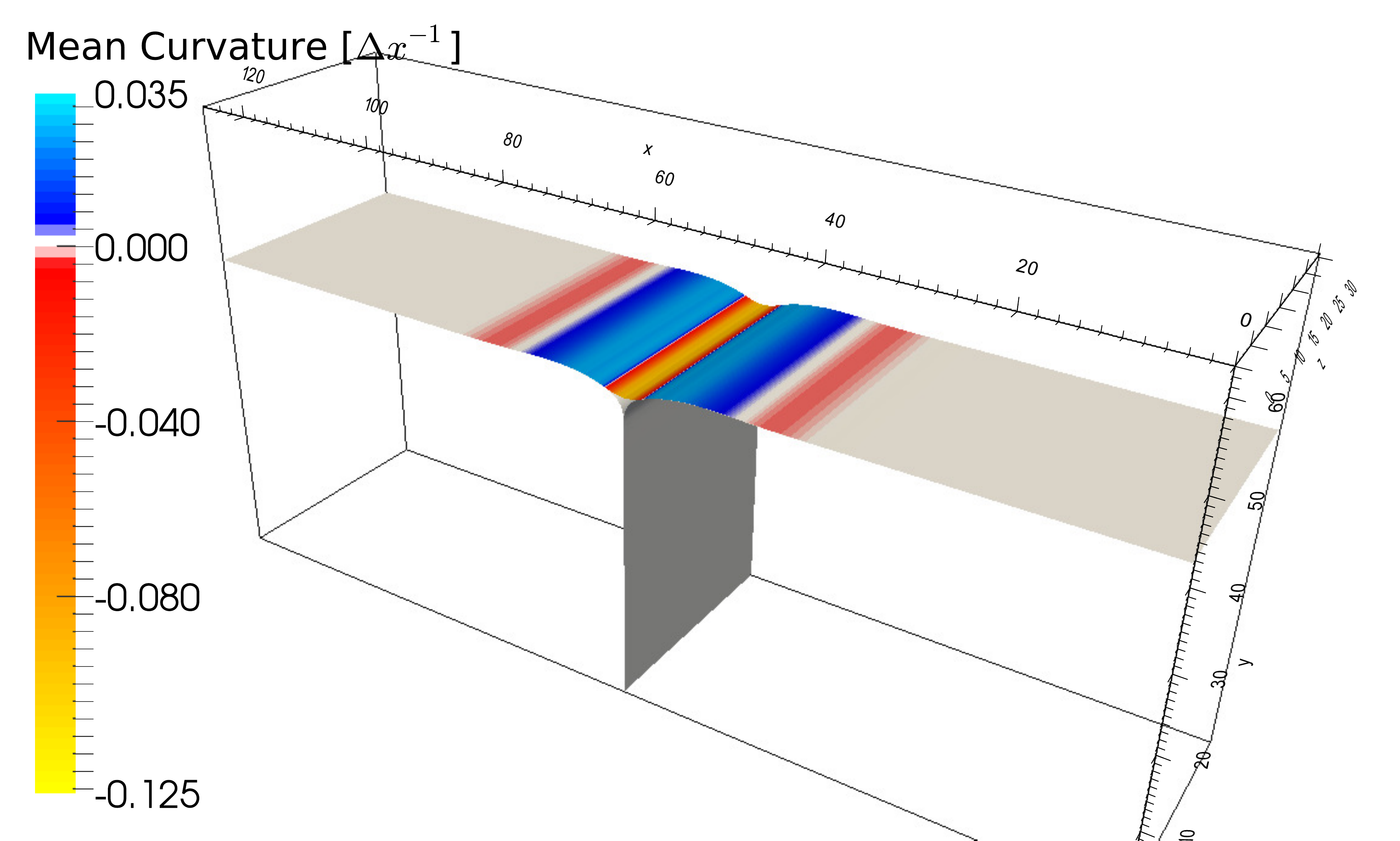}}\\
    \subfloat[\label{fig:surfaceprofile_up}] {\includestandalone[width=\textwidth,mode=buildnew]{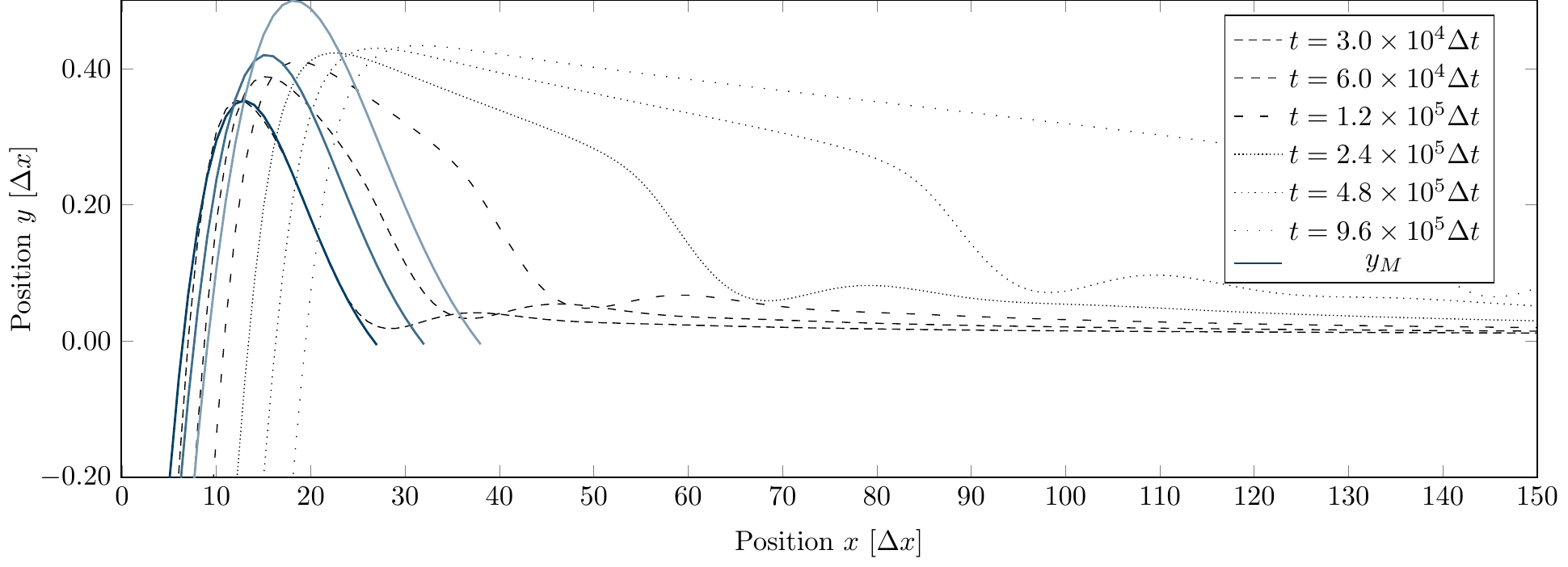}}
    \caption[Thermal grooving]{Evolution of a groove on a surface at a grain boundary by surface diffusion at
        (a) $t=0$ and (b) $t = 3 \times 10^4 \Delta t$.
        Simulation results are visualized by the isosurface of the phase-field for the gas phase $(\phi_{\text{gas}} = 0.5)$ and the two
        isosurfaces of the phase-fields representing the two solid grains
        ($\phi_\alpha = 0.5$ and $ \phi_\beta = 0.5$).
        The surface energies of both grains, $\sigma_{\text{surface}}$, and the grain boundary energy, $\sigma_{\text{GB}}$, are
        identical with $\sigma_{\text{surface}} = \sigma_{\text{GB}}= 1$ J/m$^2$.
        (c) The upscaled surface profile ($\phi_{\text{gas}} = 0.5$) with the grain boundary at $x=0$.
        The surface profile is a reconstructed phase-field contour of $\phi = 0.5$, obtained via a fourth-order polynomial and a bisection method.
        For comparison, the solid lines show the analytic solution of Mullins, Eq.~(\ref{eq:mullins}), with $m = 0.057454$ and $B = 0.00588$.
        The computation domain is discretized with (a,b) $128\times 64\times 32$ and (c) $512\times 64\times 3$ lattice nodes.
        Periodic boundary condition is applied in the $x$ and $z$-directions.
        The value of the phase-field is fixed at the boundary normal to the $y$-direction.
        Note that there is no $z$-dependence in this set-up, so that the obtained results essentially correspond to a two dimensional problem.}
    \label{fig:surfaceprofile}
\end{figure*}

In order to examine the analytically obtained results of Mullins, we consider a quasi-two dimensional set-up, where three phase-fields are initialized so that they form a flat surface with an interface perpendicular to the surface (Fig.~\ref{fig:surfaceprofile}a,b).
The evolution of the surface profile is shown in Fig.~\ref{fig:surfaceprofile_up}.
Two of the three phase-fields represent the solid grains.
The third phase-field stands for the surrounding gas phase, $\phi_{\text{gas}}$.
The surface profile is taken as the contour of the gas phase with $\phi_{\text{gas}} = 0.5$, which is reconstructed with a fourth-order polynomial and a bisection method.
This is necessary to obtain the equilibrium angles with a suitable accuracy for the following analysis.

Between the initial and equilibrium configurations, the triple junction is moving downwards, causing the interfaces to bend upwards.
Additionally, one can see that the maximum of this bended interface travels outwards.
In a private communication to Mullins~\cite{mullins_theory_1957}, Hillert pointed out that since the flow of matter immediately beyond the point of inflection is toward the origin, and since the curve has a fixed shape with enlarging size, there must be oscillations of the surface profile about the $x$-axis.
These oscillations are not predicted in the original paper by Mullins~\cite{mullins_theory_1957} who did not exclude this possibility but suspected that it would be difficult to observe these oscillations, due to the decreasing amplitude of the surface profile~\cite{mullins_theory_1957}.
The results obtained in the seminal work of Mullins, who assumes the shape invariance of the surface profile, has been confirmed by numerical solution of the underlying partial differential equations~\cite{spotz_high-order_1995,zhang_coupled_2002}.
Recently, the above mentioned non-monotonic surface profile has been reported in experiments in the case of a tungsten polycrystal~(see \cite{sachenko_observations_2004} and references therein).
As seen in Fig.~\ref{fig:surfaceprofile}, the results of the current multi-phase-field model for surface diffusion provide an independent numerical evidence for the oscillatory propagation of the surface profile during thermal grooving.
These results are also in qualitative agreement with two dimensional calculations based on a variational approach~\cite{hackl_variational_2013}.

A comparison of Eq.~(\ref{eq:mullins}) with the obtained surface profile (see Fig.~\ref{fig:surfaceprofile}) shows that Eq.~(\ref{eq:mullins}) is a good approximation for early times of the evolution and near the triple junction.
However, our simulation reveals that, the more the wave front propagates, the flatter its profile becomes in comparison to Eq.~(\ref{eq:mullins}).
A profile similar to the one obtained in this study and the profile obtained by Mullins albeit without oscillations has been experimentally observed in~\cite{gladstone_grain_2001}.

Since the oscillation pattern increases in size with time, it is more probable that they can be observed in the later stages of thermal grooving.
One can however, also not fully exclude the possible role of thermal fluctuations in damping the oscillations.
In order to investigate this aspect for the time evolution of the phase-field variable, Eq.~(\ref{eq:interface_diffusion}) shall be extended to a Langevin-type equation including the effect of thermal noise.
In addition, on the surface of a polycrystalline body which maintains multi grooving junctions, oscillations interfere with one another.
Thus, spacing between the junctions may also play an important role.
This feature is not considered in the present simulations as well.
A detailed investigation of the effects arising from thermal noise and multiple grooving junctions is left for future work.

\subsection{Von Neumann's triangle relation}
\label{sec:neumann}
In this section, we benchmark our results versus Young's law at triple junctions.
For this purpose, we take the example of thermal grooving and investigate the dependency of the equilibrium groove angle on the grain boundary energy.
Here, the equilibrium state is defined as the steady and quasi time independent profile, which establishes during late stages of thermal grooving.
In a sharp interface picture, the groove angle is defined as the angle between the left and right tangent lines of the surface profile at the triple junction.
In close similarity to this, we use the right and left  tangents of the corresponding dual phase contour lines near the triple junction.
In order to improve numerical accuracy, a high order polynomial interpolation is used to calculate the tangent lines.

Since Eq.~(\ref{eq:neumann_triangle}) is derived with the assumption of straight/planar interfaces, the accuracy will depend on the boundary conditions in the simulation box.
In order to account for this fact, the simulation of thermal grooving has been repeated for different lengths of the computation domain along the $x$-direction.

\begin{figure*}
    \subfloat[\label{fig:angle_boundary}] {\includestandalone[width=\columnwidth,mode=buildnew]{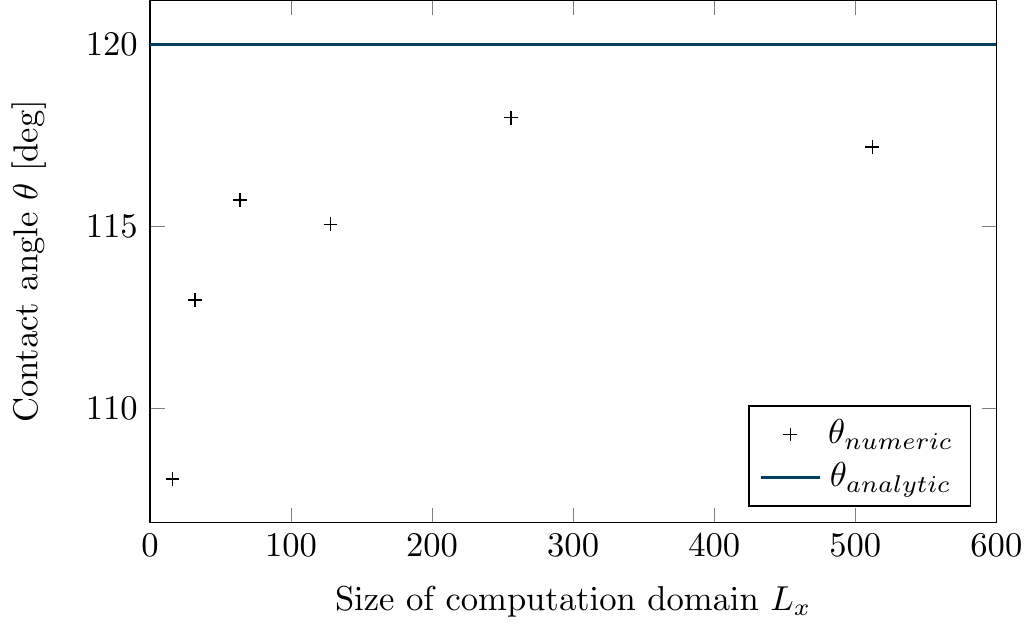}}
    \subfloat[\label{fig:angle_energy}] {\includestandalone[width=\columnwidth,mode=buildnew]{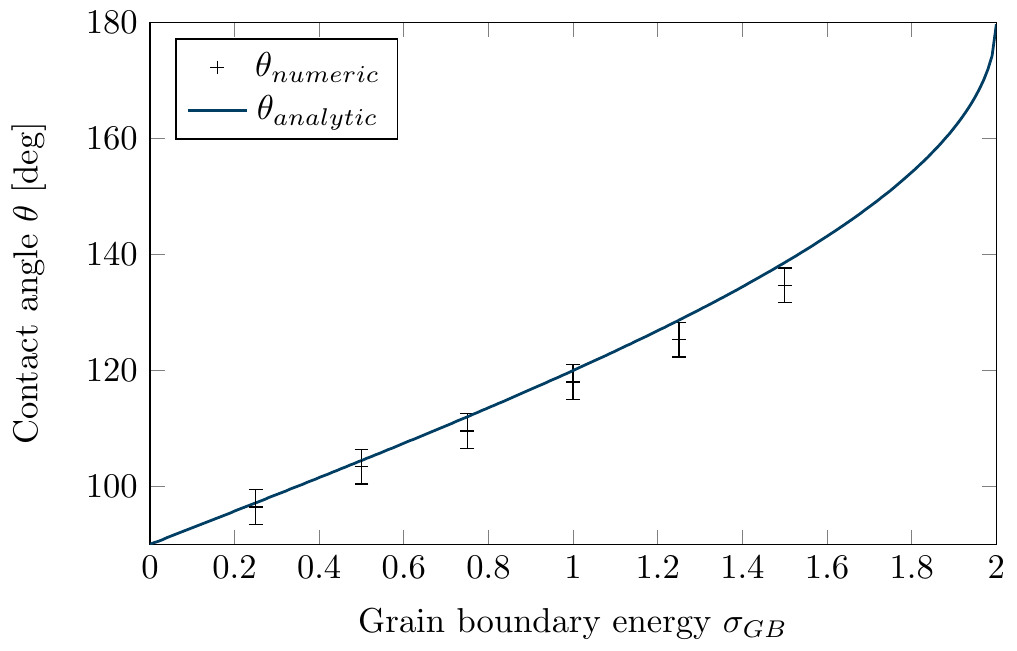}}
    \caption[Contact Angle]{The dependency of the contact angle, between the surface and the grain-boundary at a triple junction, on the size of the computation domain is shown in (a).
        The equilibrium contact angle for different grain boundary energies $\sigma_{\text{GB}}$ can be seen in (b).
        The surface energies are constant with $\sigma_{\text{surface}} = 1$J/m$^2$.
        The error bars in (b) are estimated on the basis of (a).}
\end{figure*}

We find that it is possible to obtain a relation between the size of the computation domain and the equilibrium angle (see Fig.~\ref{fig:angle_boundary}).
With this knowledge, one can minimize numerical errors and estimate the effect of the boundary.
If one assumes that the surface energy, $\sigma_{\text{surface}}$, is the same for both grains, then the groove angle, $\theta$, can be calculated using Eq.~(\ref{eq:neumann_triangle}),
\begin{dmath} \label{eq:groove_angle}
    \theta
    =
    \pi
    -
    \arctan \left(
        \frac{\sqrt{4\sigma_{\text{Surface}}^2
        - \sigma_{\text{GB}}^2}}{\sigma_{\text{GB}}}
        \right)
\end{dmath}.
The angles calculated by the simulations with the present method, and those predicted by Eq.~(\ref{eq:groove_angle}), are listed in Tab.~\ref{tab:neumann}.
The same data is plotted in Fig.~\ref{fig:angle_energy}, highlighting the close agreement between numerically obtained results and Eq.~(\ref{eq:groove_angle}).
\begin{ruledtabular}
    \begin{table}
        \begin{tabular}{c c c r r c c}
            && $\sigma_{\text{GB}}$ (J/m$^2$) & $\theta_{\text{von Neumann}}\,\, (^\circ)$ & $\theta_{\text{simulation}}\,\, (^\circ)$ && \\
            \hline
            && 0.25 &  97.2 &  96.4 && \\
            && 0.50 & 104.5 & 103.4 && \\
            && 0.75 & 112.0 & 109.6 && \\
            && 1.00 & 120.0 & 118.0 && \\
            && 1.25 & 128.7 & 125.3 && \\
            && 1.50 & 138.6 & 134.7 &&
        \end{tabular}
        \caption{The dependency of the contact angle between the surface of two grains on the grain boundary energy, $\sigma_{\text{GB}}$.
        The surface energy is kept constant at $\sigma_{\text{surface}} = 1$ J/m$^2$.
        Simulation results are listed together with the prediction of the von Neumann's relation,  Eq.~(\ref{eq:groove_angle}).}
        \label{tab:neumann}
    \end{table}
\end{ruledtabular}

\subsection{Quadruple Junctions}
\label{sec:results_quadruple}
\begin{figure*}
    \subfloat[] {\includegraphics[trim = 0 0 0 120, clip, width=0.33\linewidth]{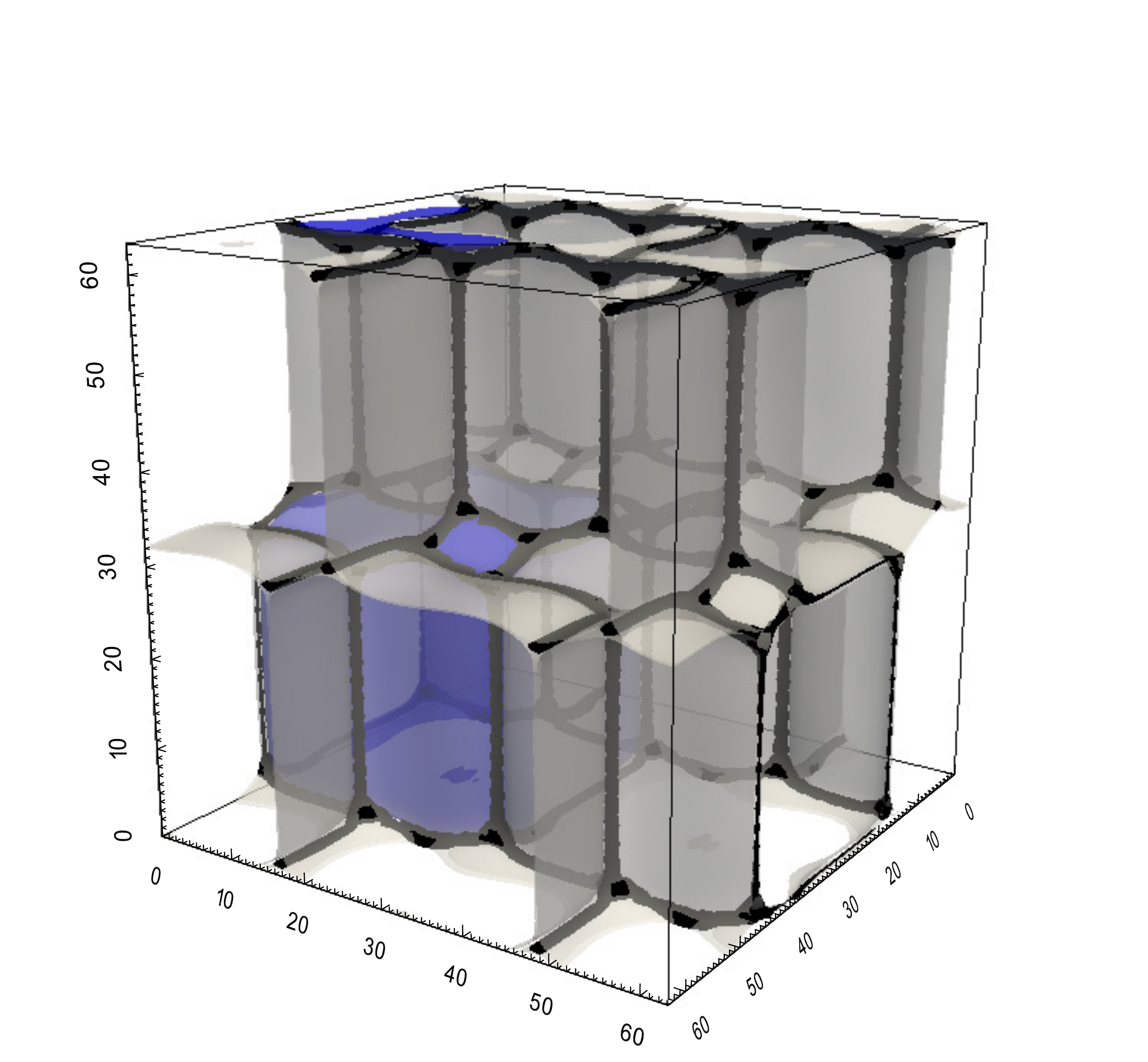}}
    \subfloat[] {\includegraphics[trim = 0 0 0 120, clip,width=0.33\linewidth]{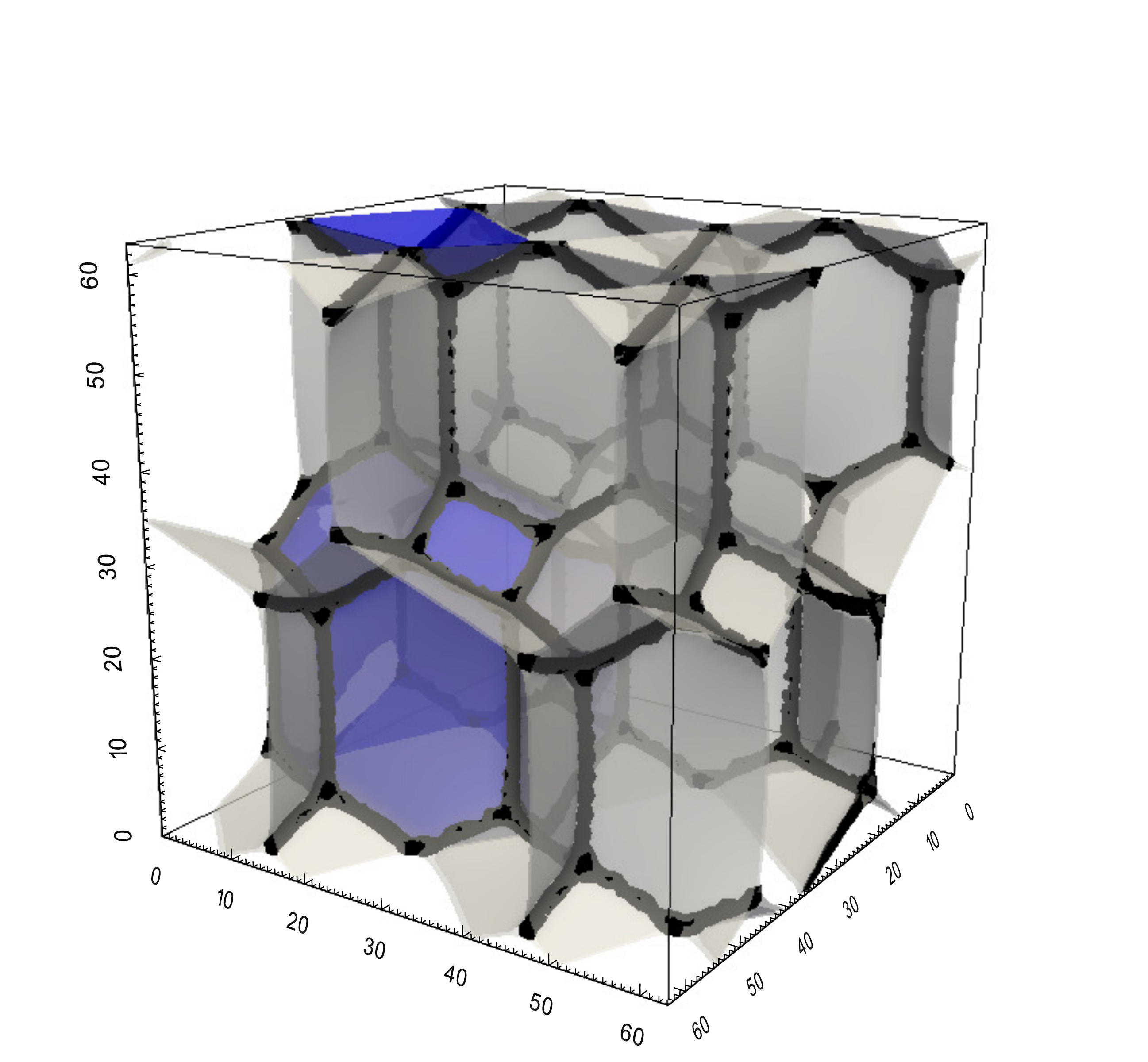}}
    \subfloat[] {\includegraphics[trim = 0 0 0 120, clip,width=0.33\linewidth]{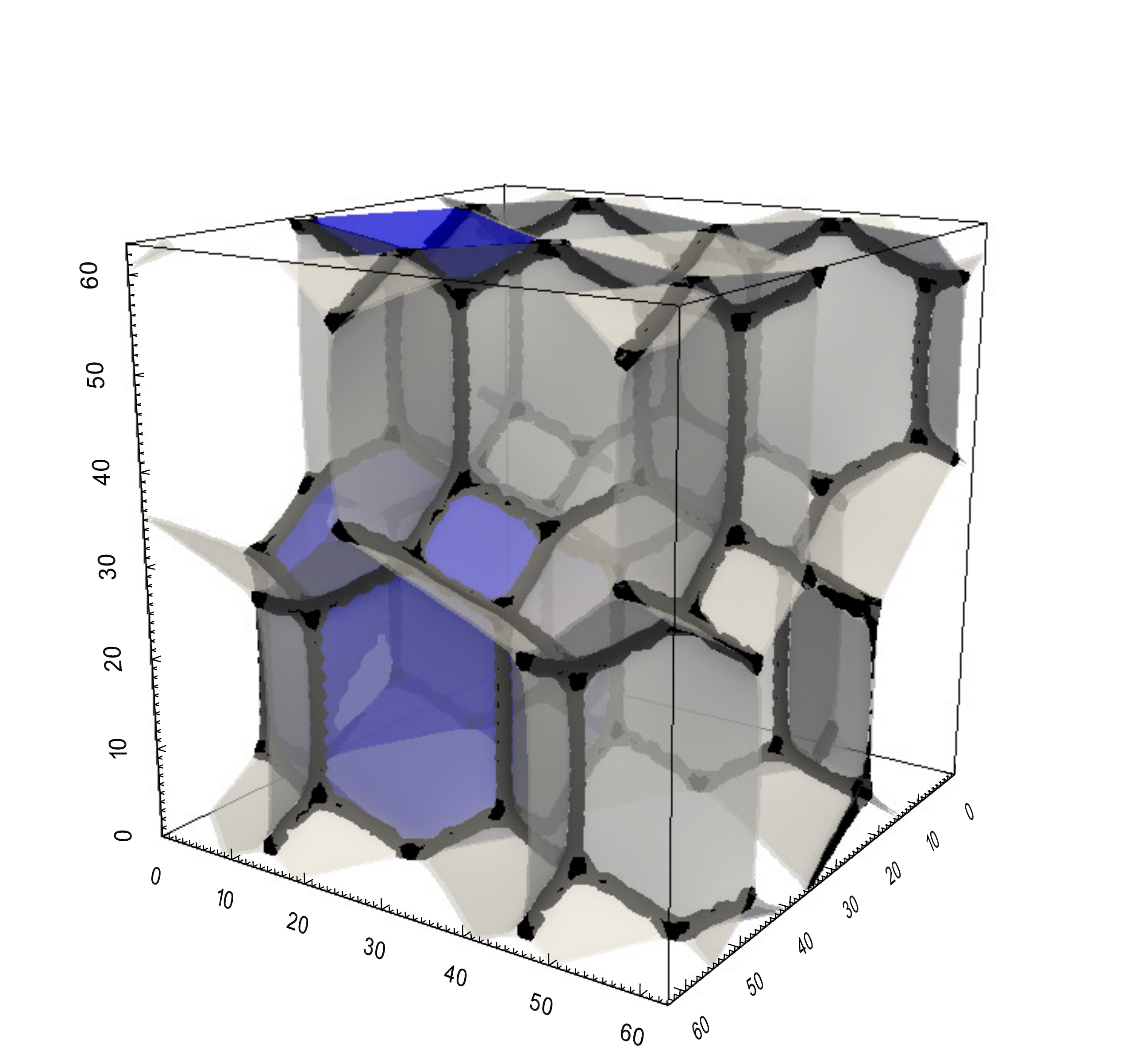} \label{fig:junction_quad_eq}}
    \caption[Quadruple junction relaxation]{The relaxation into a tessellation with truncated octahedra at (a) $t = 5 \times 10^3 \Delta t$, (b) $t = 5 \times 10^4 \Delta t$, and (c) $t = 9\times 10^5 \Delta t$.
        The computation domain is divided into eight cubic phase-fields with the same volume.
        Periodic boundary condition is used along all the coordinate directions $x$, $y$ and $z$.
        The cubic phases are shifted in a way that there are only triple lines and quadruple junctions.
        The dual interfaces are shown with transparent greyish planes and the triple lines with black lines.
        Additionally, one phase is highlighted with less transparent planes.
        The domain is discretized into $64\times 64\times 64$ lattice nodes.}
    \label{fig:junction_quad}
\end{figure*}

Although the above simulations of thermal grooving are performed in three dimensions, they are essentially equivalent to a two dimensional situation due to the cylindrical symmetry of the set-up (no $z$-dependence).
For multi-grain (multi-phase) systems in 3D, however, vertex points (quadruple and higher order junctions) may also exist in which more than three grains meet.
In order to investigate this type of scenario, a multi-grain set-up is constructed by dividing the simulation domain into eight cubes such that only triple lines and quadruple points form between the grains (see Fig.~\ref{fig:junction_quad}a).
By surface minimization, one can see that the grains arrange themselves into truncated octahedra (see Fig.~\ref{fig:junction_quad}c).
The equilibrium angle between the triple lines at a quadruple junction is either $90^\circ$, when the lines are connected by a square plane, or $120^\circ$, when the lines are connected by a hexagonal plane.
We find that the dihedral angle at a triple line between two hexagonal planes of truncated octahedron is roughly $109^\circ$ and the angle between a hexagonal and a square plane is $\approx 125^\circ$.
These are well in line with the values of $109^\circ 28^' 16^"$ and $125^\circ 15^' 51^"$ expected from geometrical consideration.
However, from a generalization of the von Neumann's triangle relation to quadruple junctions, one would expect that the system does not form truncated octahedra and that all dihedral angles have the same value.
The difference between the simulation results and the prediction of a generalized von Neumann's triangle relation for quadruple junctions is probably due to the conservation constraint in the periodic set-up of simulations.

It is quite simple, however, to show that the sum of forces which act alongside the triple lines on a quadruple point is zero (see App.~\ref{app:force_okta}).

\subsection[Droplets On a Surface]{Droplets on a Deformable Surface}
\label{sec:results_droplets}
An important application of the phase-boundary diffusion model proposed in this work is the simulation of nano-clusters on free surfaces.
In particular, the model allows to study the entrenching of nano-clusters, reported in recent experiments \cite{kohler_thermally_2013}.

\begin{figure*}
    \subfloat[]{\includestandalone[width=0.33\textwidth, mode=buildnew]{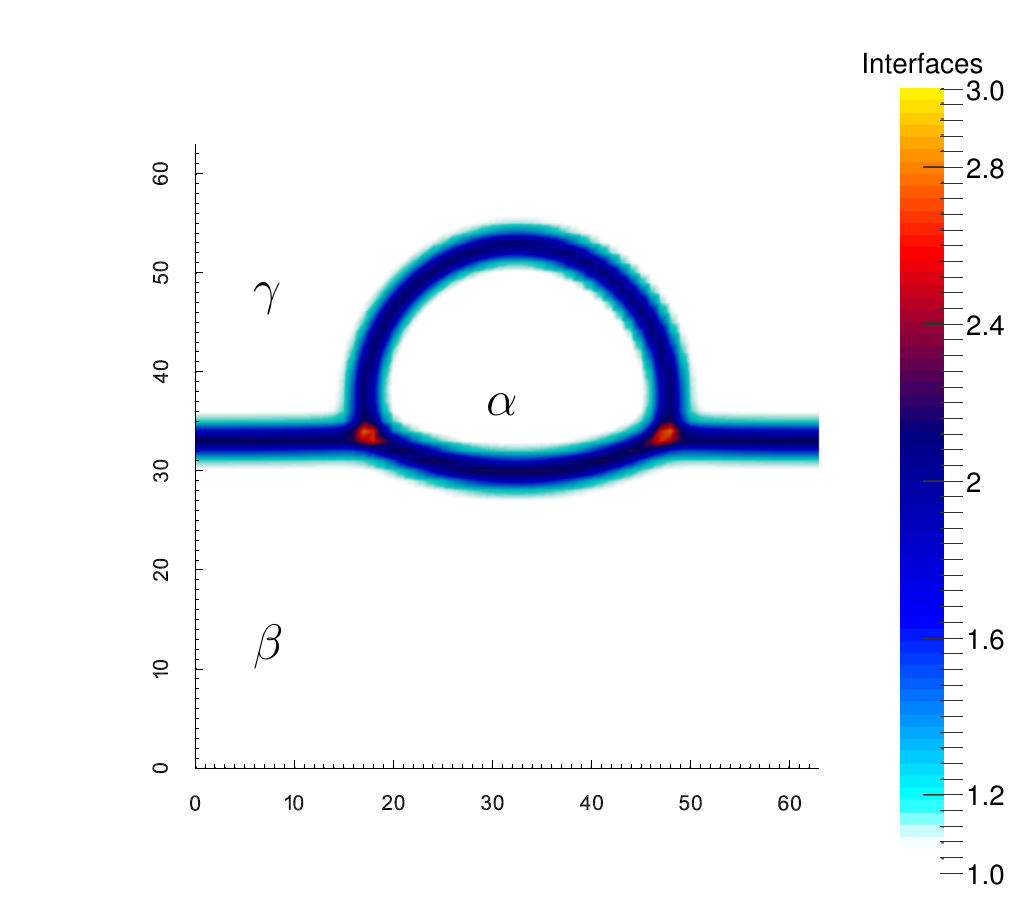}}
    \subfloat[]{\includestandalone[width=0.33\textwidth, mode=buildnew]{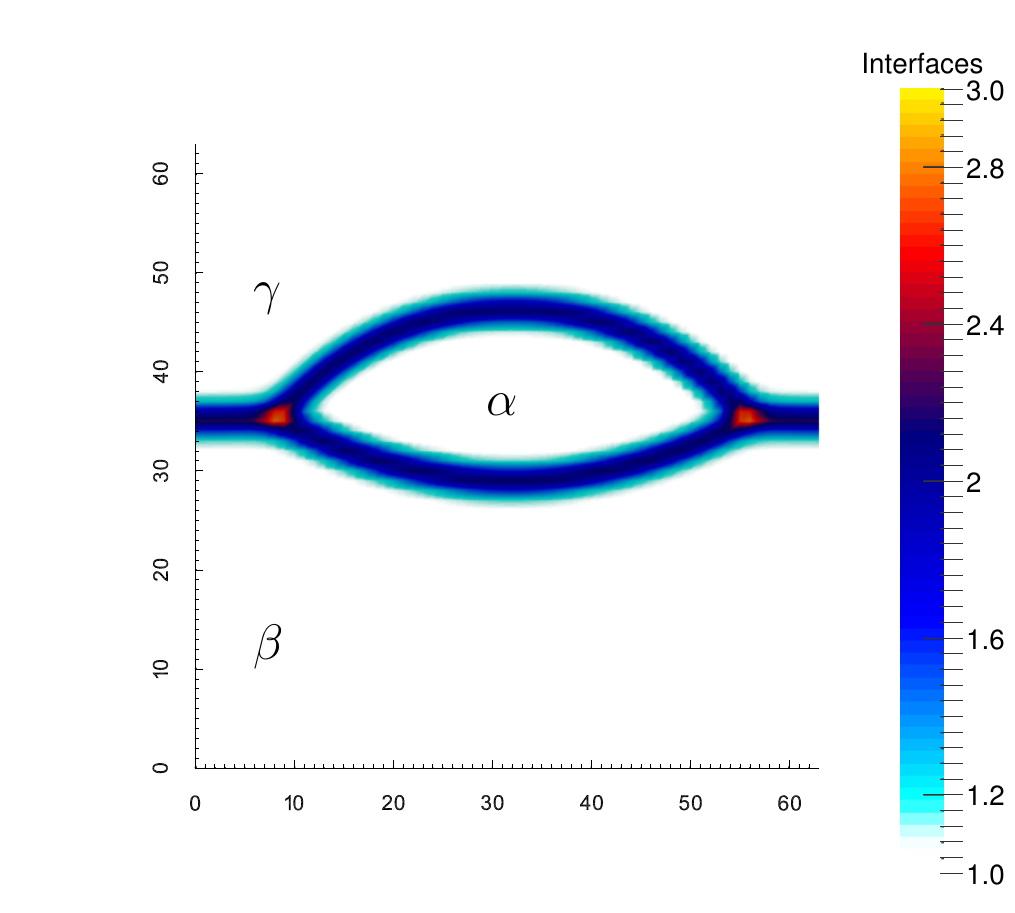}}
    \subfloat[\label{fig:IEwetting}]{\includestandalone[width=0.33\textwidth, mode=buildnew]{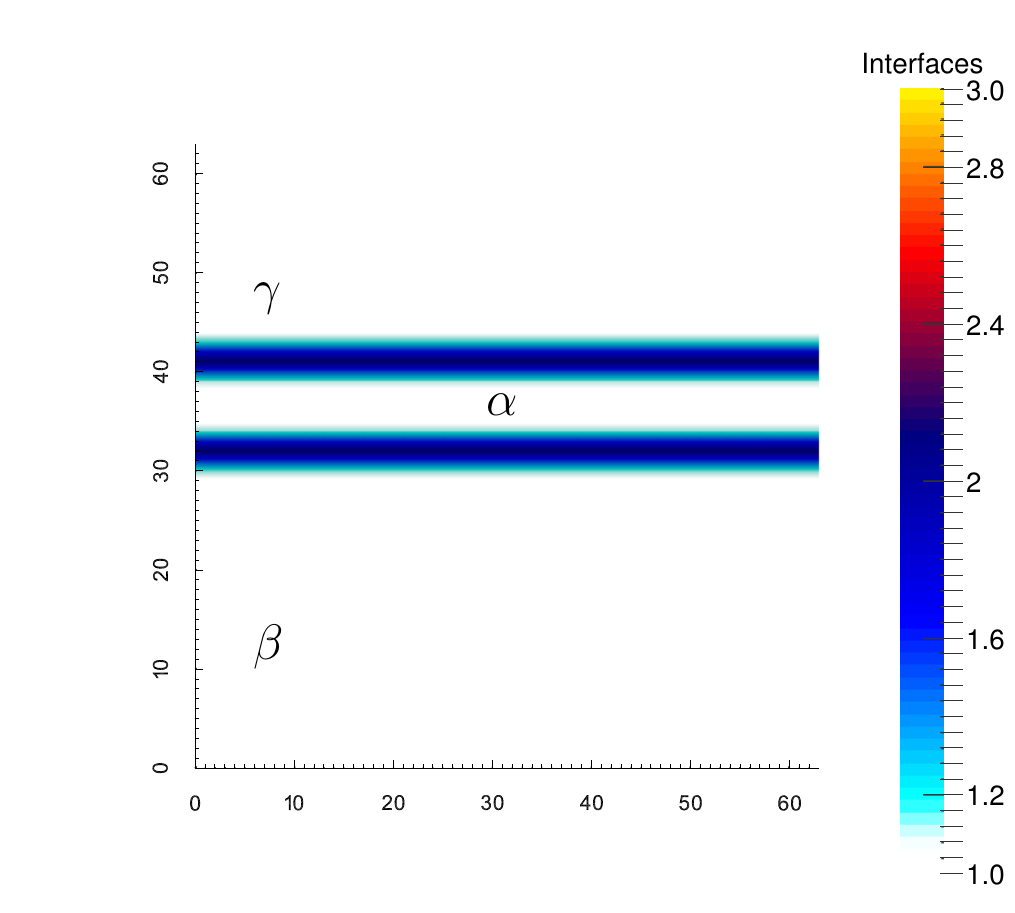}}
    \caption{The configuration of multiple solid droplets on a deformable surface
        after $t =  10^6 \Delta t$ is shown for different interface energies with
        (a) $\sigma_{\alpha \gamma} = 0.5$ J/m$^2$, $\sigma_{\beta  \gamma} = 0.75$ J/m$^2$, and $\sigma_{\alpha \beta}  = 1$ J/m$^2$;
        (b) $\sigma_{\alpha \gamma} = 0.5$ J/m$^2$, $\sigma_{\beta  \gamma} = 1$ J/m$^2$, and $\sigma_{\alpha \beta}  = 0.75$ J/m$^2$;
        and (c) $\sigma_{\alpha \gamma} = 0.5$ J/m$^2$, $\sigma_{\beta  \gamma} = 10$ J/m$^2$, and $\sigma_{\alpha \beta}  = 0.5$ J/m$^2$.
        The computation domain is discretized by $64 \times 64$ lattice nodes, periodic boundary condition in $x$-direction and fixed boundary conditions in $y$-direction are applied.}
    \label{fig:interface}
\end{figure*}

In order to study the influence of the interface energy on the equilibrium contact angles between the nano-clusters and the free surface, we have placed a spherical droplet on a surface and observed its wetting dynamics.
The equilibrium configurations for different interface energies are shown in Fig.~\ref{fig:interface}.
A complete wetting of the surface, beyond Eq.~(\ref{eq:neumann_triangle}), is visible in Fig.~\ref{fig:IEwetting}.

\begin{figure*}
    \subfloat[] {\includegraphics[width=0.45\linewidth] {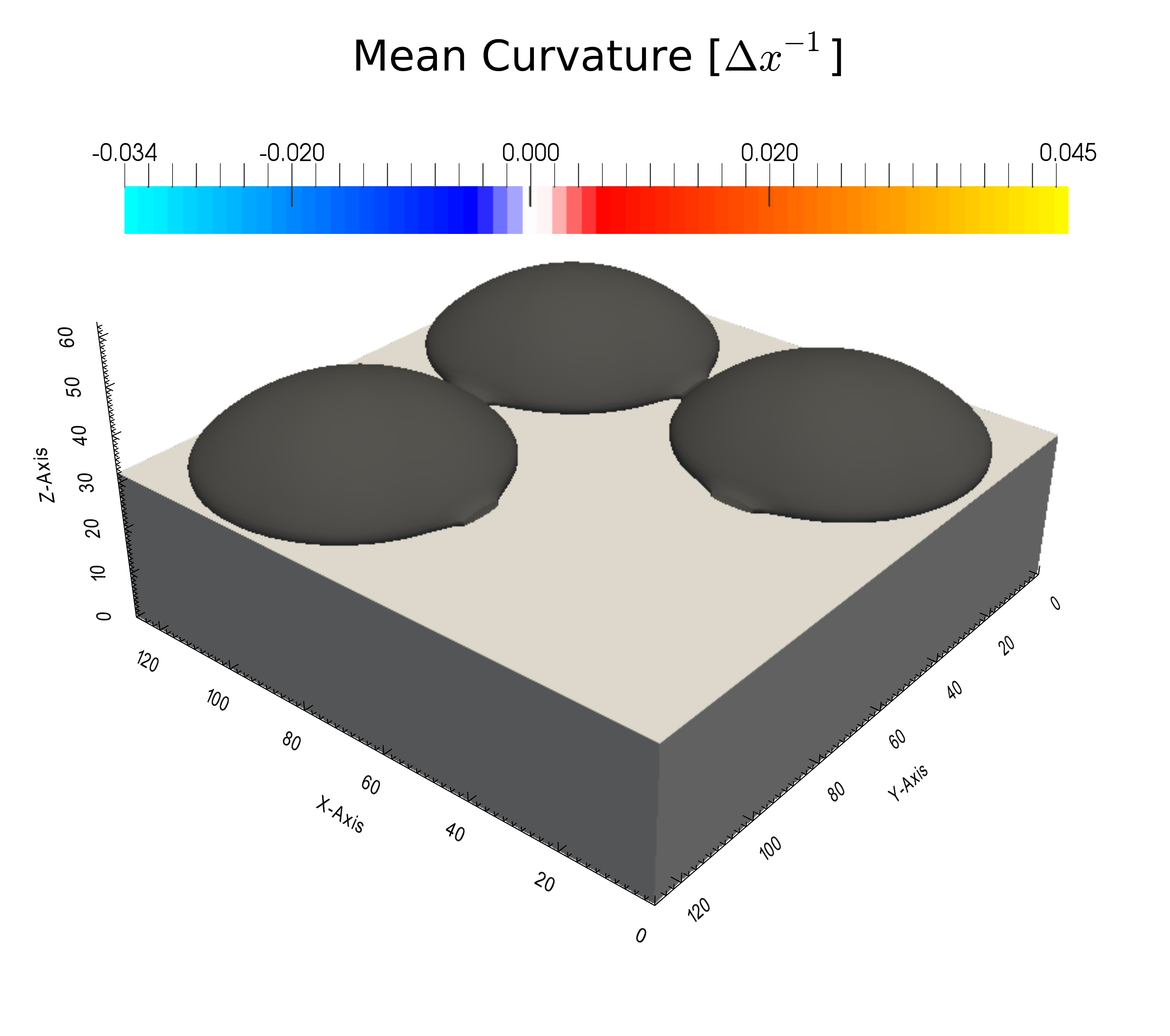}}
    \subfloat[\label{eq:perimeter}] {\includegraphics[width=0.45\linewidth]{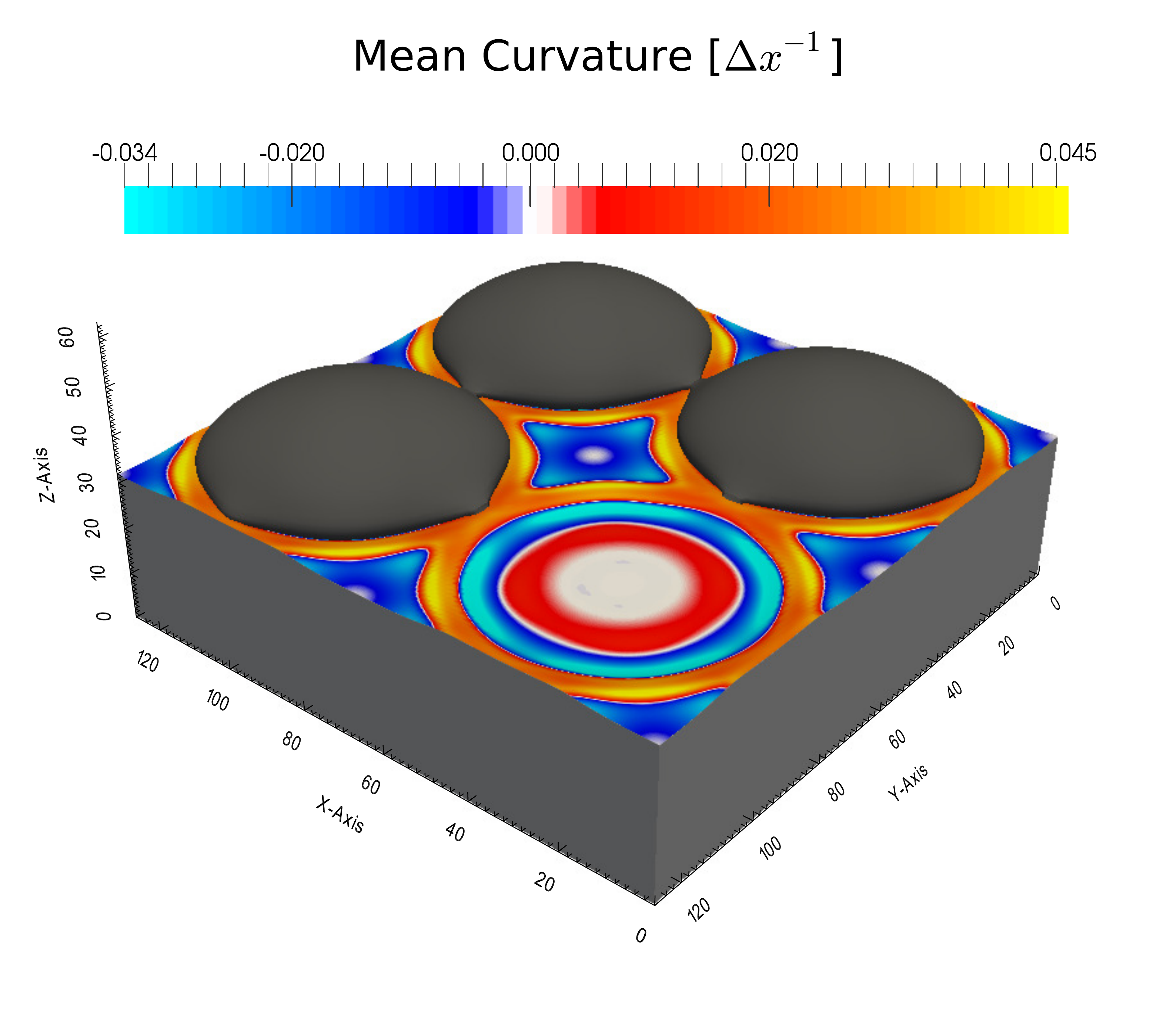}}

    \subfloat[] {\includegraphics[width=0.45\linewidth]{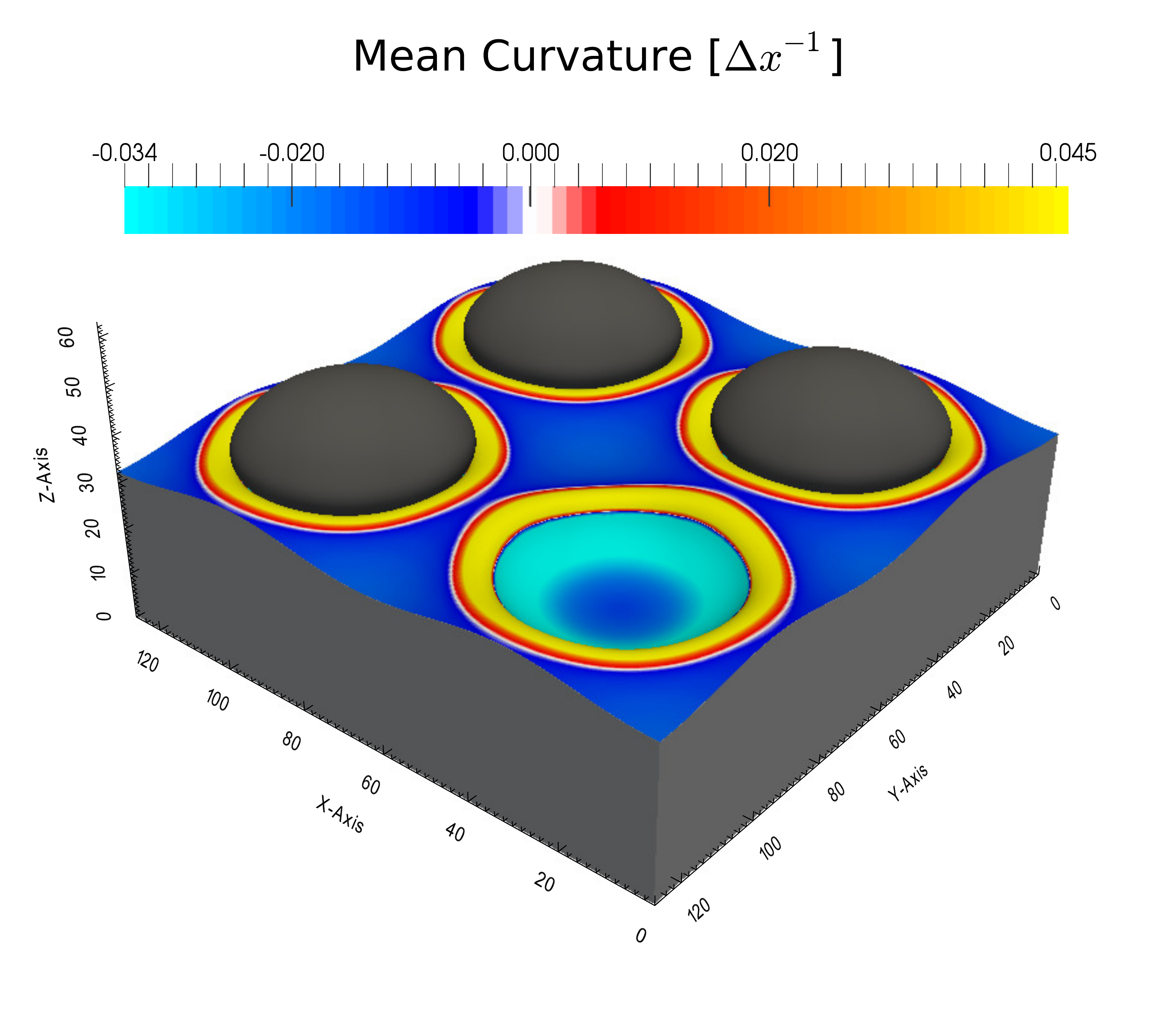}}

    \caption[Multiple droplets on a surface]{The entrenching of multiple nano-clusters $\alpha$ into a deformable surface $\beta$ by surface/phase-boundary diffusion with (a) $t=0$, (b) $t = 5 \times 10^3 \, \Delta t$, and (c) $t = 1.5 \times 10^5 \, \Delta t$.
        The interface energies are set to $\sigma_{\alpha \beta} = 1.4$ J/m$^2$, $\sigma_{\alpha \gamma} = 1.4$ J/m$^2$ and $\sigma_{\beta \gamma} = 2$ J/m$^2$  ($\gamma$ is the phase surrounding the phases $\alpha$ and $\beta$).
        One of the four clusters has been removed in the plot in order to show the base area which is colored according to its curvature.
        The simulation domain is discretized into $128\times 128 \times 64$ grid cells.}
    \label{fig:nanosclusters}
\end{figure*}

The equilibrium configurations shown in Fig.~\ref{fig:interface} do not resemble the geometry of the nano-clusters observed in~\cite{kohler_thermally_2013} where elevated perimeters have been observed which surround the `crater-like' entrenchments of the nano-clusters.
This perimeter can be explained as intermediate state of the entrenching process, in a sense that the entire system is not in equilibrium.
This interpretation is confirmed by our simulations, shown in Fig.~\ref{fig:nanosclusters}, where four initially connected droplets on surface are simulated with identical interface energies.
One can clearly see that the droplets separate from each other and entrench into the surface.
The perimeter is visualized in Fig.~\ref{eq:perimeter}.
Thus, the present model is not only able to reproduce experimentally observed results but also allows to uncover the transient character of the elevated perimeter in the entrenching process.
A full investigation of the multi-nano-clusters on free surfaces is in progress.

\section{Summary}
\label{sec:summary}
A conserved multi-phase-field model is proposed to describe the physical effect of surface/phase-boundary diffusion.
Starting from the non-conserved multi-phase-field model~\cite{steinbach_phase-field_2009}, a pairwise continuity equation is proposed for temporal evolution of the surface/phase-boundaries by curvature-driven diffusion.

The model is applied to investigate the dynamics of the surface profile during thermal grooving.
Simulation results indicate that the shape invariant solution of Mullins~\cite{mullins_theory_1957} is a good approximation to the early stages of the thermal grooving process.
Moreover, new evidence is provided for the oscillations of the surface profile, first proposed by Hillert.
Additional simulations of thermal grooving at variable interface energies are then performed in order to gain further confidence on the reliability of the present approach.
This is achieved by a check of the late time profiles obtained from these simulations against the von Neumann's triangle relation for equilibrium angles at a triple junction.

As a forecast on future applications of the proposed model, the annealing dynamics of nano-clusters on initially flat surfaces is investigated.
The reported crater-like structure with an elevated perimeter~\cite{kohler_thermally_2013} is found to be a transient non-equilibrium state during nano-cluster annealing, thus shedding light onto this complex process from a dynamic perspective.
The proposed multi-phase-field method is generic and can be used to study any type of surface and phase-boundary phenomena which contains conserved fields.

\appendix

\section{Static Equilibrium Solution}
\label{app:travel}

In the case of two phase-fields with
$ \phi_\alpha = 1 - \phi_\beta = \phi $,
the generalized diffusion potential introduced in Eq.~(\ref{eq:generalized_chemical_potential}) simplifies to
\begin{dmath} \label{eq:simple_potiential}
    \psi
    =
    \frac{8 \sigma_{\alpha\beta} }{\eta}
    \left[
        \frac{\eta^2}{\pi^2}
        \nabla^2 \phi
        +
        2 \phi - 1
    \right]
\end{dmath},
where we assumed $\phi \left(1-\phi\right) \in \left[0,1\right]$ so that the
partial derivate of the obstacle potential (2nd term in Eq.~(\ref{eq:Iab_b})) reduces to
$ \partial \left\vert\phi\left(1-\phi\right) \right \vert/\partial \phi =  2 \phi - 1 $.

Using this information, one finds that the equilibrium solution of Eq.~(\ref{eq:interface_diffusion}) for a planar interface, situated at $x=0$ between two phases,  is identical to Eq.~(\ref{eq:travel_sphere}) with the difference that the radial distance $r$ shall be replaced by the Cartesian coordinate along the direction normal to the interface, say $x$.

\section[Forces at quadruple junction]{Forces at a quadruple junction of a tessellation with truncated octahedra}
\label{app:force_okta}

If one considers a quadruple in a tessellation with truncated octahedra as in Fig.~\ref{fig:junction_quad_eq}, one sees that four triple lines are connected to that quadruple junction.
The configuration of the phase-fields at each of these triple lines is the same.
So it is reasonable to assume that the magnitude, $m$, of the forces alongside the triple line is the same and only their orientation differs.
These forces can be written as follows:
\begin{dgroup}
    \begin{dmath}
        \vec{F_1} = \frac{m}{\sqrt 2} \left(-1, -1,  0 \right)^T
    \end{dmath}
    \begin{dmath}
        \vec{F_2} = \frac{m}{\sqrt 2} \left( 1, -1,  0 \right)^T
    \end{dmath}
    \begin{dmath}
        \vec{F_3} = \frac{m}{\sqrt 2} \left( 0,  1, -1 \right)^T
    \end{dmath}
    \begin{dmath}
        \vec{F_4} = \frac{m}{\sqrt 2} \left( 0,  1, 1 \right)^T
    \end{dmath}.
\end{dgroup}
One can easily check that the sum of $\vec F_1$, $\vec F_2$, $\vec F_3$ and $\vec F_4$ is zero, and the angles between the forces are either $90^\circ$ or $120^\circ$.
\begin{acknowledgments}
    Financial support by the German Research Foundation DFG within Priority Program SPP1713 under the grants DA 1655/1-1, STE 1116/20-1 and the DFG-project VA 205/17-1 is gratefully acknowledged.
\end{acknowledgments}

\bibliography{ms.bib}

\end{document}